\documentclass[runningheads]{llncs}
\usepackage{graphicx}
\usepackage{xcolor}
\usepackage{underscore}
\usepackage[backref]{hyperref}
\usepackage{xcolor}
\usepackage{comment}
\usepackage{amssymb}
\usepackage{tabularray}

\hypersetup{
	colorlinks,
	linkcolor={blue!80!black},
	citecolor={blue!70!black},
	urlcolor={blue!80!black}
}

\begin{document}
 \title{Fast Medical Shape Reconstruction via Meta-learned Implicit Neural Representations}

\titlerunning{Fast Medical Shape Reconstruction via Meta-learned INRs}

\author{Gaia Romana De Paolis\inst{1} \and
Dimitrios Lenis\inst{1} \and 
Johannes Novotny\inst{1} \and
Maria Wimmer \inst{1} \and 
Astrid Berg \inst{1} \and
Theresa Neubauer \inst{1} \and 
Philip Matthias Winter\inst{1} \and
David Major \inst{1} \and
Ariharasudhan Muthusami \inst{1} \and
Gerald Schröcker \inst{2} \and
Martin Mienkina  \inst{2} \and
Katja Bühler \inst{1}
}

\authorrunning{G.R. De Paolis et al.}
%
\institute{VRVis Zentrum für Virtual Reality und Visualisierung Forschungs-GmbH, \\ Vienna, Austria \\
\email{depaolis@vrvis.at} \and
GE Healthcare Austria GmbH, Zipf, Austria}
\maketitle              

\begin{abstract}
Efficient and fast reconstruction of anatomical structures
plays a crucial role in clinical practice. Minimizing retrieval and processing times not only potentially enhances swift response and decision-making in critical scenarios but also supports interactive surgical planning and navigation. Recent methods attempt to solve the medical shape reconstruction problem by utilizing implicit neural functions. However, their performance suffers in terms of generalization and computation time, a critical metric for real-time applications. To address these challenges, we propose to leverage meta-learning to improve the network parameters initialization, reducing inference time by an order of magnitude while maintaining high accuracy. We evaluate our approach on three public datasets covering different anatomical shapes and modalities, namely CT and MRI. Our experimental results show that our model can handle various input configurations, such as sparse slices with different orientations and spacings. Additionally, we demonstrate that our method exhibits strong transferable capabilities in generalizing to shape domains unobserved at training time.
	
\end{abstract}
%
%
\begin{figure}[t] 
	\centering
	\includegraphics[width=\textwidth]{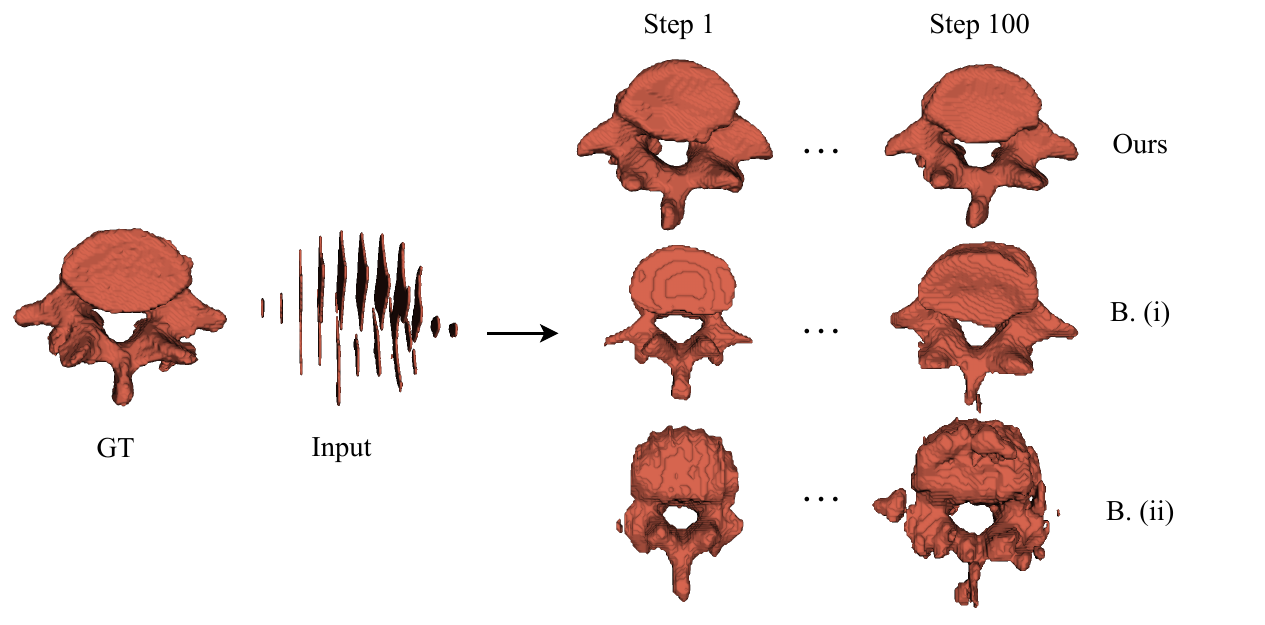}
	\caption{Our model based on meta-learned INRs can reconstruct anatomical shapes in a single optimization step, achieving accuracy comparable to both baselines (B.~(i), B.~(ii)), while also being significantly faster at inference time.}
	\label{fig:vertebra}
	
\end{figure}

\section{Introduction}
Fast and accurate 3D medical shape representation is crucial for critical healthcare tasks as time-critical diagnosis, computer-aided interactive surgical planning and image-guided interventions \cite{ambellanStatisticalShapeModels2019,benyedderDeepLearningBiomedical2021,chenShapeRegistrationLearned2021}. This is notably evident in navigation systems, as in spinal or cardiac surgery that require high-resolution real-time visualization of the target shapes \cite{wilsonImageGuidedNavigationSpine2024,chenShapeRegistrationLearned2021,banerjeeAutomated3DWholeHeart2022}. In practice, the acquired measurements used to obtain a shape representation can vary in the level of detail captured, according to the imaging modality and specific medical application. Numerous scenarios demand to minimize the number of measurements needed for image acquisition, thereby losing resolution. Particularly, in CT imaging, sparse sampling reduces radiation exposure for patients, while in MRI, it accelerates the scanning procedure, reducing the risk of motion artifacts and patient discomfort \cite{reedDynamicCTReconstruction2021,shenNeRPImplicitNeural2022}.
As a consequence, shapes derived from manual or (semi-)automatic segmentations of those scans only provide a sparse representation of the actual 3D shape of an anatomical object \cite{sanderReconstructionCompletionHighresolution2023a}. In this context, medical shape reconstruction aims to recover detailed anatomical structures from limited or incomplete segmentations.
Many approaches have studied this ill-posed problem \cite{benyedderDeepLearningBiomedical2021,molaeiImplicitNeuralRepresentation2023}, however, there is a lack of studies focusing on both fast and highly accurate 3D reconstructions in the medical field. This gap limits the applicability of shape reconstruction methods in real-time applications e.g. surgical guidance and navigation \cite{chenShapeRegistrationLearned2021} where time complexity is an important metric in the method selection \cite{benyedderDeepLearningBiomedical2021}. 

In this work, we propose to combine meta-learning and Implicit Neural Representation (INR) functions to swiftly reconstruct 3D anatomical shapes from sparse segmentations of scans with a limited set of observations (Fig. \ref{fig:vertebra}).
Existing methods for surface meshes are commonly based on statistical shape models, which typically necessitate a dense point correspondence between training shapes. Such correspondences are difficult to obtain from sparse segmentation masks \cite{heimannStatisticalShapeModels2009,tothovaProbabilistic3DSurface2020a,amiranashviliLearningShapeReconstruction2022}. Moreover, they require extracting a mesh from voxel-based segmentations, which is generally computationally intensive and hard, especially for incomplete segmentations \cite{amiranashviliLearningShapeReconstruction2022}.
In voxel space, the dominant class of shape reconstruction methods successfully uses Convolutional Neural Networks (CNNs) to learn mapping functions from sparse representations to complete images, typically by leveraging super-resolution methods and generative models \cite{shenPatientspecificReconstructionVolumetric2019,cerrolaza3DFetalSkull2018,turellaHighResolutionSegmentationLumbar2021,lyuMRISuperResolutionEnsemble2020}. Despite their efficacy, these methods are limited by their need for large training datasets, their instability to structural changes and their inability to generalize across image modalities or anatomical sites \cite{molaeiImplicitNeuralRepresentation2023}.
Furthermore, they typically encode input signals following an explicit approach, by discretizing the input space into separate elements (e.g. point clouds, meshes, voxel grids). This limits their ability to process volumetric grids with varying spacings and resolutions \cite{tewariAdvancesNeuralRendering2022a,meschederOccupancyNetworksLearning2019,molaeiImplicitNeuralRepresentation2023}.
Recent works \cite{chibaneImplicitFunctionsFeature2020a,parkDeepSDFLearningContinuous2019,meschederOccupancyNetworksLearning2019,yeGIFSNeuralImplicit2022,yingAdaptiveLocalBasis2023} address these shortcomings by employing INR functions as an alternative approach for
shape representation and completion.
Within this context, an INR is a Multi-Layer Perceptron (MLP) that takes a spatial 3D coordinate as input and predicts the corresponding intensity value.
The main advantage is that INRs operate in a continuous domain, enabling them to represent an image or object at arbitrary resolutions, independently from the voxel space. Consequently, in contrast to discrete representations, INRs can be significantly more memory efficient while preserving fine details \cite{tewariAdvancesNeuralRendering2022a,molaeiImplicitNeuralRepresentation2023,sitzmannImplicitNeuralRepresentations2020a}.  
To this end, previous works have investigated the concept of learning and utilizing a prior over the distribution of shapes of a given structure of interest \cite{meschederOccupancyNetworksLearning2019,sitzmannImplicitNeuralRepresentations2020a,parkDeepSDFLearningContinuous2019}. Predominant approaches in natural or scene representation presuppose a low-dimensional latent space where the embedding is decoded into a function through the use of hypernetworks \cite{sitzmannImplicitNeuralRepresentations2020a,sitzmannSceneRepresentationNetworks2020} or via concatenation-based conditioning \cite{parkDeepSDFLearningContinuous2019,meschederOccupancyNetworksLearning2019}.
In the medical domain, Shen et al. \cite{shenNeRPImplicitNeural2022} propose NeRP framework that integrates INRs with reconstruction from sparsely sampled medical images without the need of any training data, by leveraging a previous scan of the same patient.
While their study demonstrates the effectiveness of combining prior embedding and implicit neural representations, it still requires tens of optimization steps for each subject. Additionally, in terms of generalization, it necessitates a specific prior training for each patient and scan.
Amiranashvili et al. \cite{amiranashviliLearningShapeReconstruction2022} adopt the auto-decoder scheme, demonstrating that implicit functions are able to perform high-resolution shape reconstruction by learning a shape prior from anisotropic volumetric segmentations.
However, during inference, it needs multiple optimization steps to specialize to a new shape, which may take several seconds per volume.
Addressing inference speed, recent studies \cite{sitzmannMetaSDFMetalearningSigned2020,tancikLearnedInitializationsOptimizing2021,dupontDataFunctaYour2022} apply meta-learning algorithms to learn the initial weights for INR networks, leading to faster convergence in several tasks, including 2D CT reconstruction \cite{tancikLearnedInitializationsOptimizing2021}.  

Inspired by the aforementioned works, we introduce a meta-learning based method to further improve the reconstruction performance for 3D medical shapes in terms of inference time efficiency and generalization, when only partial observations of the medical image are available. 
The main contributions of this work can be summarized as follows:

\begin{itemize}
	\item To the best of our knowledge, this is the first study that recovers complete 3D anatomical shapes from sparse measurements via meta-learning.   
	\item Our model is able to reconstruct new, unseen anatomical shapes in just one optimization step, achieving accuracy on the level of a state-of-the-art approach and standard gradient optimization, while being significantly faster during inference.
	\item Experimental results on three publicly available datasets demonstrate our model generalizability across various input configurations, such as sparse slices with different orientations and spacings. Moreover, we show that our method has strong transferable capabilities in generalizing to shape domains that differ from the prior domain and are not observed during training. 

\end{itemize}

\section{Methods}
\subsection{Problem Formulation}
Our objective is to quickly recover a complete 3D anatomical shape, given a set of observations that sparsely covers its spatial extent.
Formally, we assume that we are given a dataset $\mathcal{D}$ of $N \in \mathbb{N}$ medical volume images $V$ with corresponding segmentations $S$, i.e.  $\mathcal{D} = {\left\{(V_{i}, S_{i})\right\}}_{i=1}^{N}$. 
In this study, we interpret a segmentation $S_i$ as the shape to be reconstructed from a set of partial observations of $V_i$. $S_i$, with $i = 1,..., N$, is characterized by a regular grid points with each point represented by spatial coordinates $\mathbf{x}$ and a corresponding ground-truth segmentation value $z : ( \mathbf{x} \in  \mathbb{R}^3 ) \rightarrow \{0, 1\}$.
Following Amiranashvili et al. \cite{amiranashviliLearningShapeReconstruction2022}, we simulate anisotropic voxel size by constraining the available observations of $V_i$ to lay on slices, sampled with distance \textit{w}.
We model the 3D shape as a neural implicit representation by a network $f_{\theta}$, that defines the boundary of the object \cite{meschederOccupancyNetworksLearning2019}.
We attempt to learn a shared prior over all available subjects in $\mathcal{D}$ and then use this initialization to reconstruct a new, unseen shape from partial observations.
We search for the initialization $\theta_0^*$ that (a) allows fast convergence (b) serves as a strong starting point for gradient descent, when optimizing a new signal within the manifold characterized by $\mathcal{D}$.
\noindent
\textbf{Architecture}
The architecture is an MLP that takes 3D coordinates $\mathbf{x}$ as input and returns the probability of this point being inside the related shape \cite{meschederOccupancyNetworksLearning2019}, that is $f_{\theta} : ( \mathbf{x} \in  \mathbb{R}^3 ) \rightarrow [0, 1]$, where $\theta$ represents the weights of the MLP. 
We apply periodic activation functions \cite{sitzmannImplicitNeuralRepresentations2020a}, in order to address the complicated nature of anatomical shapes.

\subsection{Fast Shape Reconstruction via Meta-Learning}  \label{subsec:shapeprior}
To obtain the final reconstruction, our method is split into two phases:
\begin{enumerate}
 \item \textit{Shape Prior Meta-Learning}, where a meta-learning algorithm is used to learn a shared prior $\theta_0^*$ over all available subjects in $\mathcal{D}$.
 \item \textit{Shape Reconstruction}, where, given the shape prior as initialization, the network can be optimized to reconstruct any shape, in the given distribution, using only the available sparse measurements as input. 
\end{enumerate} 
\subsubsection{Shape Prior Meta-Learning } 
For our purpose, meta-learning can be formalized in the context of few-shot learning where the goal is to learn a model that can rapidly adapt to new tasks \cite{sitzmannMetaSDFMetalearningSigned2020,tancikLearnedInitializationsOptimizing2021,hospedalesMetaLearningNeuralNetworks2021}. In this study, we view representation and reconstruction of a shape as a dedicated task.
Building upon the formulation proposed by Sitzmann et al. \cite{sitzmannMetaSDFMetalearningSigned2020}, we sample, in the forward pass, \textit{context} and \textit{target} observations from $S_i$, denoted as $S_i^{c} \in S_i$ and $S_i^{t} \in S_i$.
\textit{Meta-learning} employs two learning algorithms, commonly named \textit{inner} and \textit{outer} loop \cite{finnModelAgnosticMetaLearningFast2017}, that, here, aim to learn the representations of $S_i^{c}$ and $S_i^{t}$. At each iteration, the inner algorithm learns a new task $S_i^{c}$ i.e. the shape to be reconstructed, which is subsequently used as initialization for the outer loop to learn the target task. We leverage MAML \cite{finnModelAgnosticMetaLearningFast2017}, an optimization-based meta-learning algorithm in which the model learns new tasks with few steps of gradient descent. 
The inner algorithm (Eq.~\ref{eq:innerloop}) computes the weight values $\theta^L$, with $L \in \mathbb{N}$ optimization steps.
In the outer loop (Eq.~\ref{eq:outerloop}), we use the learned $\theta^{L}$ to predict the occupancy probability on the target set $S^{t}$ and to learn the initial weights $\theta_0$. Inner and outer loop can be formalized as follows:
\begin{equation} \label{eq:innerloop}
	\theta^{l+1} = \theta^{l} - \alpha\nabla \sum_{(\mathbf{x}, z(\mathbf{x}))\in S_i^{c}} \mathcal{L}(f_{\theta^l}(\mathbf{x}), z(\mathbf{x}))
\end{equation}
\begin{equation}\label{eq:outerloop}
	\theta_0^{p+1} = \theta_0^{p} - \beta\nabla \sum_{(\mathbf{x}, z(\mathbf{x}))\in S_i^{t}} \mathcal{L}(f_{\theta^L}(\mathbf{x}), z(\mathbf{x}))
\end{equation} 
where $\nabla$ is the gradient, $l \in \left(0, L\right]$ and $p \in \left(0, P\right]$ are generic optimization steps, $\alpha$ and  $\beta$ are the step sizes for inner and outer loop respectively. $f_{\theta^l}$ and $f_{\theta^L}$ are the network evaluated with parameters $\theta^l$, $\theta^L$ at steps $l$, $L$ respectively.

We chose a common loss function for both loops, since both involve fitting a volumetric representation. It is described as:

\begin{equation} \label{eq:loss}
	\mathcal{L} = \mathcal{L}_{BCE}(f_\theta(\mathbf{x}), z(\mathbf{x})) + \mathcal{L}_{Dice}(f_\theta(\mathbf{x}), z(\mathbf{x})) + \lambda \frac{1}{q} \|\theta\|_2^2
\end{equation}

where $\mathcal{L}_{BCE}$ is the Binary Cross-Entropy (BCE) and $\mathcal{L}_{Dice}$ is the Dice loss component. $q$ is the MLP's number of weights and $\lambda$ is a weighting parameter.
Since representations for recovering $S_i$ are not unique, we finally include a regularization on $\theta$, as proposed by Stizmann et al.~\cite{sitzmannImplicitNeuralRepresentations2020a}.

After optimization, the final $\theta_0^*$ = $\theta_0^P$, encodes the information from the prior available representations and can be used as initialization for the optimization of a new, unseen segmentation.

\subsubsection{Shape Reconstruction}
We can reconstruct the full shape of an unseen target volume from sparse observations by initializing the model parameters with the learned prior $\theta_0^*$. The reconstruction is obtained through two steps.

First, we optimize the network to learn the neural representation of the new target volume measurements. The network $f_\theta$ is sampled at the available positions of voxels from $S$. This optimization process follows the same scheme as described in the previous section.
Subsequently, we generate the final reconstruction by evaluating the trained network across all the spatial coordinates in the volume space. 

\section{Experimental Setup} \label{sec:baseline}
\subsubsection{Datasets}
We conducted experiments on three publicly available datasets, that cover several varieties of medical data, including two modalities i.e. CT, MRI, different sizes and anatomical shapes with different complexity.
The first dataset, Verse '19 \cite{sekuboyinaVerSeVertebraeLabelling2021}, contains 3D CT segmentations of vertebrae. Following the setup described in \cite{amiranashviliLearningShapeReconstruction2022}, we extract subvolumes of size $ 128 \times 128 \times 128$ with an isotropic spacing of $1$ $mm^3$ around each vertebra, obtaining 287 volumes. 
The other two datasets are from the Medical Segmentation Decathlon \cite{antonelliMedicalSegmentationDecathlon2022} and consist of 281 3D CT pancreas and 20 3D MRI heart segmentations.
The pancreas and heart volumes are cropped to $256 \times 128 \times 64$ and  $ 128 \times 128 \times 128$ size respectively, ensuring that the entire target shapes are encompassed.
In the following, we name $\mathcal{D}_v$, $\mathcal{D}_p$ and $\mathcal{D}_h$, respectively for vertebra, pancreas and heart datasets.

\subsubsection{Baselines}
We compare our results with two established methods in medical field, referred to as B.~(i) and B.~(ii), that investigate the use of shape prior knowledge for reconstruction from sparse measurements of medical data.

B.~(i) is the method from Amiranashvili et al. \cite{amiranashviliLearningShapeReconstruction2022} that is the current state-of-the-art in shape reconstruction from sparse segmentations. It adopts the auto-decoder setup in which the MLP is conditioned on a latent vector corresponding to each single subject, enabling generalization across a set of shapes. At test time, it performs a search to optimize the latent vector for a new subject, that can require several seconds per volume.

B.~(ii) is inspired by NeRP \cite{shenNeRPImplicitNeural2022}. Here, we compare our meta-learned initialization with the one derived from pre-training the MLP on the training set, using a standard gradient optimization.

\subsubsection{Implementation Details}
To generate the training data, we sample every 8-th slice along the sagittal axis, skipping slices in between so to simulate realistic sparse segmentations, as outlined in \cite{amiranashviliLearningShapeReconstruction2022}.
All experiments are implemented employing an MLP consisting of 7 linear layers with 128 hidden units and sinusoidal activation function \cite{sitzmannImplicitNeuralRepresentations2020a}. We use MAML \cite{finnModelAgnosticMetaLearningFast2017} as meta-learning algorithm over 5 inner-loop update steps and we initialize $\alpha$ as $1 \times 10 ^ {-5}$. 
Our model and B.~(ii) are optimized using the ADAM optimizer \cite{kingmaAdamMethodStochastic2017}, with a learning rate of $1 \times 10 ^ {-5}$ and loss function described in Eq.~\ref{eq:loss}, where $\lambda = 1 \times 10 ^ 2$. We train our model and B.~(ii) for 2500 epochs.
To train B.~(i), we use the same training setup described by the authors in \cite{amiranashviliLearningShapeReconstruction2022}. The inference was performed on an NVIDIA A100 80GB GPU for all models.

\subsubsection{Evaluation Setup}

The objective of our experiments is to demonstrate that our framework expedites the reconstruction process while maintaining accuracy on the level of baseline approaches. Therefore, we evaluate our model and the baselines comparing both speed and accuracy at different optimization steps.
In order to assess the generalizability of our approach, we consider several reconstruction tasks as sparse slices with different orientations and spacings. We also provide evaluations on target shapes outside the prior domain to explore the transfer learning capabilities of our model to other target shapes, an application not addressed in previous works \cite{amiranashviliLearningShapeReconstruction2022,shenNeRPImplicitNeural2022}.
All experiments are evaluated using 5-fold cross validation.
Ground-truth segmentations were solely utilized for comparing predictions in the inference phase and were not used for training purposes.
The quantitative comparison between reconstruction and ground-truth is conducted using Dice Similarity Coefficient (\textit{\textit{DSC}}) score.

\section{Results}
\subsection{Fast Convergence}

\begin{table}[t]

	\caption{Average number of iterations and corresponding seconds necessary to match the Dice score ($DSC_1$) reached by our method in one optimization step, for vertebra ($\mathcal{D}_v$), pancreas ($\mathcal{D}_p$) and heart ($\mathcal{D}_h$). B.~(i) and B.~(ii) are described in Sec.~\ref{sec:baseline}.
	}
	
	\begin{tabular}{lcclcclcc}
		\cline{1-3} \cline{5-6} \cline{8-9}
		& \multicolumn{2}{c}{\textbf{$\mathcal{D}_v$}}                                                                                                             &  & \multicolumn{2}{c}{\textbf{$\mathcal{D}_p$}}                                                                                                        &  & \multicolumn{2}{c}{\textbf{$\mathcal{D}_h$}}                                                                                                           \\ \cline{1-3} \cline{5-6} \cline{8-9} 
		\multicolumn{1}{c}{\textbf{}} & \textbf{\begin{tabular}[c]{@{}c@{}}$\#$ iters \\ to $DSC_1$\end{tabular}} & \textbf{\begin{tabular}[c]{@{}c@{}}Rec.\\ time (s)\end{tabular}} &  & \textbf{\begin{tabular}[c]{@{}c@{}}$\#$ iters \\ to $DSC_1$\end{tabular}} & \textbf{\begin{tabular}[c]{@{}c@{}}Rec.\\ time (s)\end{tabular}} &  & \textbf{\begin{tabular}[c]{@{}c@{}}$\#$ iters \\ to $DSC_1$\end{tabular}} & \textbf{\begin{tabular}[c]{@{}c@{}}Rec.\\ time (s)\end{tabular}} \\ \cline{1-3} \cline{5-6} \cline{8-9} 
		B. (i)                        & 26.73 $\pm$ 16.7                                                             & 0.68 $\pm$ 0.4                                                           &  & 44.43 $\pm$ 19.3                                                             & 1.45 $\pm$ 0.6                                                      &  & 33.85 $\pm$ 9.3                                                              & 1.12 $\pm$ 0.3                                                      \\ \cline{1-3} \cline{5-6} \cline{8-9} 
		B. (ii)                       & $>$ 150                                                                   & $>$  5                                                                &  & $>$ 150                                                                   & $>$  5                                                           &  & $>$ 150                                                                   & $>$  5                                                           \\ \cline{1-3} \cline{5-6} \cline{8-9} 
		Ours                          & 1                                                                         & \textbf{0.09 $\pm$ 0.1}                                                  &  & 1                                                                         & \textbf{0.08 $\pm$ 0.05}                                            &  & 1                                                                         & \textbf{0.11 $\pm$ 0.18}                                            \\ \cline{1-3} \cline{5-6} \cline{8-9} 
	\end{tabular}
		\label{tab:timing}
\end{table}

 \begin{figure}[t]
	\centering
	\includegraphics[width=.3\textwidth]{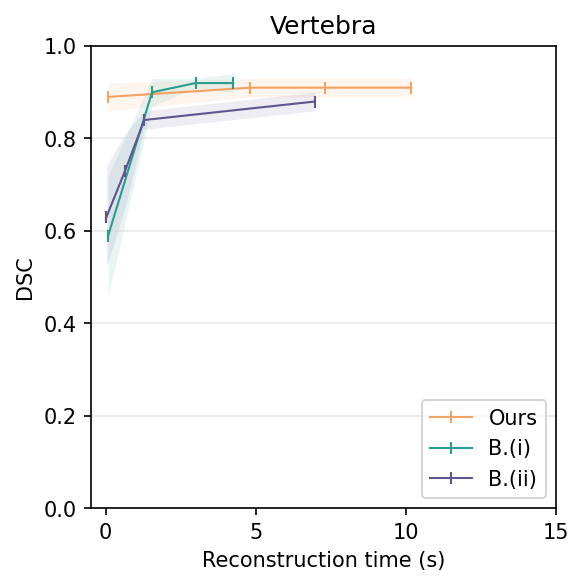}
	\includegraphics[width=.3\textwidth]{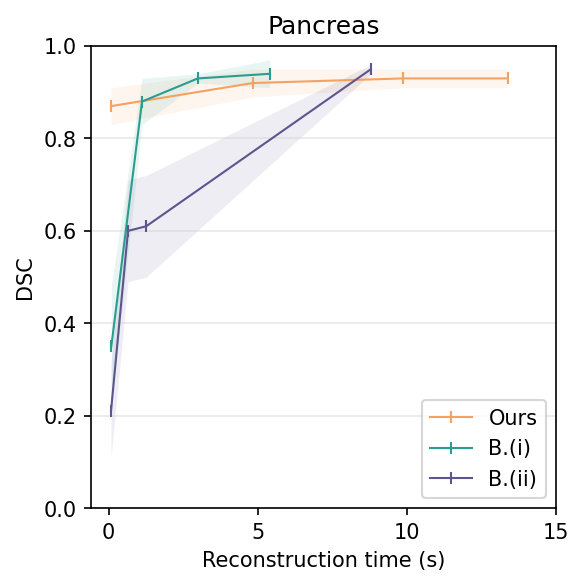}
	\includegraphics[width=.3\textwidth]{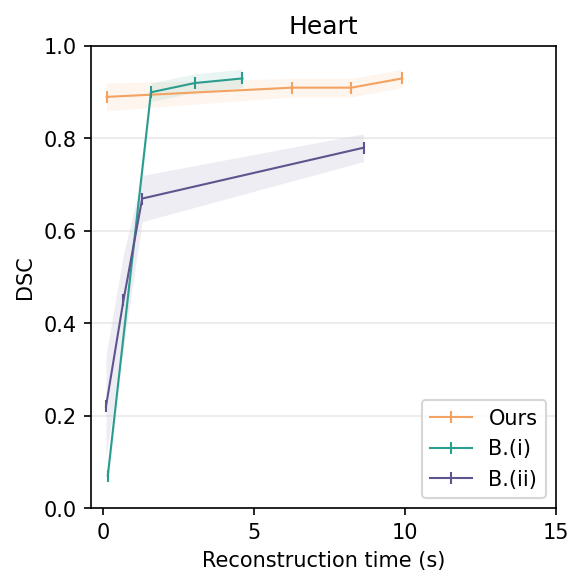}
	\caption{Average $DSC$ over reconstruction time measured at step 1, 50, 100 and at convergence. The shaded areas represent the corresponding standard deviations. The meta-learned initialization (ours) allows to fast ($\sim 0.1s$) recover a new sample, with accuracy on the level of B.(i) and B.(ii) at convergence.}
	\label{fig:graph}
\end{figure}

\subsubsection{Reconstruction from Sagittal Slices} \label{sec:resultSagittal} 
We evaluate our method on the same reconstruction configuration used for training. 
Table~\ref{tab:steps} presents $DSC$ values at 1, 50 and 100 optimization steps, including performance at convergence. 
We extend our comparisons to results achieved after more steps as baselines in related studies utilize longer iterations (1000-2000) \cite{amiranashviliLearningShapeReconstruction2022,shenNeRPImplicitNeural2022}.
Convergence values are determined with an early stopping patience of 10 steps. The number of iterations and corresponding optimization time required for convergence are detailed in Table~\ref{tab:convergence} in the Appendix. 
After a single optimization step (Table~\ref{tab:steps}), our method considerably outperforms both alternative approaches on all tested datasets.
Table~\ref{tab:timing} reports the number of iterations and seconds required to achieve the \textit{DSC} value that our model reaches in a single iteration, i.e. $DSC_1$.
Considering this target score, our method is one order of magnitude faster ($\sim0.1s$) than both baselines, making it suitable for real-time applications.
At convergence, our model perform on par with B.~(i) and B.~(ii) overall (Table~\ref{tab:steps}, Fig.~\ref{fig:graph}).
Nevertheless,  to achieve $DSC$ values (e.g. $0.89 \pm 0.03$ for vertebra) in a single step ($\sim0.1s$) comparable to baselines' performance at convergence (e.g. $0.92 \pm 0.02$ in $4.25 \pm 1.0 s$ by B.(i) and $0.88 \pm 0.02$ in $6.97 \pm 1.3 s$ by B.(ii) for vertebra). This demonstrates our model provide an effective compromise between high-quality and fast reconstruction.  
This trend is further illustrated in the graphs (Fig.~\ref{fig:graph}), that depict the average $DSC$ over reconstruction time, confirming that our model reaches high accuracy significantly faster than the compared methods, on all tested shapes.
 
 \begin{table}[t]
	\centering
	\caption{Average $DSC$ for reconstructions from input segmentations with $w = 8$ along sagittal axis, after 1 ($DSC_1$), 50 ($DSC_{50}$), 100 ($DSC_{100}$) optimization steps and at convergence ($DSC_{conv}$).}
	\begin{tabular}{llcllc}
		\cline{2-6}
		& \multicolumn{1}{c}{\textbf{}} & $DSC_{1}$                & \multicolumn{1}{c}{$DSC_{50}$}      & \multicolumn{1}{c}{$DSC_{100}$} & $DSC_{conv}$             \\ \cline{2-6} 
		\multicolumn{1}{c}{\textbf{$\mathcal{D}_v$}} & B.(i)                         & 0.59 $\pm$ 0.13          & \multicolumn{1}{c}{0.90 $\pm$ 0.03} & \textbf{0.92 $\pm$ 0.01}        & \textbf{0.92 $\pm$ 0.02} \\ \cline{2-6} 
		& B.(ii)                        & 0.63 $\pm$ 0.11          & 0.74 $\pm$ 0.08                     & 0.84 $\pm$ 0.03                 & 0.88 $\pm$ 0.02          \\ \cline{2-6} 
		& Ours                          & \textbf{0.89 $\pm$ 0.03} & \textbf{0.91 $\pm$ 0.02}            & 0.91 $\pm$ 0.02                 & 0.91 $\pm$ 0.02          \\ \cline{2-6} 
		\multicolumn{1}{c}{\textbf{$\mathcal{D}_p$}} & B.(i)                         & 0.35 $\pm$ 0.13          & 0.88 $\pm$ 0.05                     & \textbf{0.93 $\pm$ 0.01}        & 0.94 $\pm$ 0.03          \\ \cline{2-6} 
		& B.(ii)                        & 0.21 $\pm$ 0.10          & 0.60 $\pm$ 0.11                     & 0.61 $\pm$ 0.11                 & \textbf{0.95 $\pm$ 0.01} \\ \cline{2-6} 
		& Ours                          & \textbf{0.87 $\pm$ 0.04} & \textbf{0.92 $\pm$ 0.03}            & 0.93 $\pm$ 0.02                 & 0.93 $\pm$ 0.02          \\ \cline{2-6} 
		\multicolumn{1}{c}{\textbf{$\mathcal{D}_h$}} & B.(i)                         & 0.07 $\pm$ 0.02          & 0.90 $\pm$ 0.02                     & 0.92 $\pm$ 0.02                 & \textbf{0.93 $\pm$ 0.02} \\ \cline{2-6} 
		& B.(ii)                        & 0.22 $\pm$ 0.11          & 0.45 $\pm$ 0.10                     & 0.67 $\pm$ 0.05                 & 0.78 $\pm$ 0.03          \\ \cline{2-6} 
		& Ours                          & \textbf{0.89 $\pm$ 0.03} & \textbf{0.91 $\pm$ 0.02}            & \textbf{0.93 $\pm$ 0.02}        & 0.91 $\pm$ 0.02          \\ \cline{2-6} 
	\end{tabular}
	\label{tab:steps}
\end{table}

  \begin{table}[t]

	\caption{Average DSC for reconstructions from input segmentations with $w = 16$ (a) along sagittal axis and with $w = 8$ along coronal and axial axis (b), after 1 ($DSC_1$) and 100 ($DSC_{100}$) optimization steps.}
	\begin{minipage}{0.45\textwidth} \label{tab:sagittal}
		\centering
		\begin{tabular}{llcl}
			\cline{2-4}
			& \multicolumn{1}{c}{\textbf{}} & $DSC_1$                   & \multicolumn{1}{c}{$DSC_{100}$}        \\ \cline{2-4} 
			\cline{2-4} 
			\multicolumn{1}{c}{\textbf{$D_v$}} & B.~(i)                  & 0.59 $\pm$ 0.13          & \textbf{0.90 $\pm$ 0.03}             \\ \cline{2-4} 
			& B.~(ii)                 & 0.63 $\pm$  0.11         & \multicolumn{1}{c}{0.76 $\pm$  0.06} \\ \cline{2-4} 
			& Ours                          & \textbf{0.82 $\pm$ 0.03} & 0.83 $\pm$ 0.03                      \\ \cline{2-4} 
			\multicolumn{1}{c}{\textbf{$D_p$}} & B.~(i)                  & 0.34 $\pm$ 0.14          & \textbf{0.91 $\pm$ 0.03}             \\ \cline{2-4} 
			& B.~(ii)                 & 0.43 $\pm$ 0.13          & 0.66 $\pm$ 0.10                      \\ \cline{2-4} 
			& Ours                          & \textbf{0.85 $\pm$ 0.04} & 0.90 $\pm$ 0.03                      \\ \cline{2-4} 
			\multicolumn{1}{c}{\textbf{$D_h$}}    & B.~(i)                  & 0.07 $\pm$ 0.02          & \textbf{0.91 $\pm$ 0.02}             \\ \cline{2-4} 
			& B.~(ii)                 & 0.22 $\pm$ 0.11          & 0.51 $\pm$ 0.06                      \\ \cline{2-4} 
			& Ours                          & \textbf{0.83 $\pm$ 0.04} & \multicolumn{1}{c}{0.84 $\pm$ 0.04}  \\ \cline{2-4} 
		\end{tabular}\par\vspace{0.5em}
		(a) 
	\end{minipage}
	\begin{minipage}{0.45\textwidth}
		\centering

		\begin{tabular}{lllll}
			\hline                                                                                   
			\multicolumn{1}{c}{}                  & \multicolumn{2}{c}{\textit{\textbf{Coronal}}}              & \multicolumn{2}{c}{\textit{\textbf{Axial}}}                \\ \hline
			& \multicolumn{1}{c}{$DSC_1$} & \multicolumn{1}{c}{$DSC_{100}$} & \multicolumn{1}{c}{$DSC_1$} & \multicolumn{1}{c}{$DSC_{100}$} \\ \hline
			\textbf{$\mathcal{D}_v$}                                  & 0.89 $\pm$ 0.0      & 0.91 $\pm$ 0.1         & 0.88 $\pm$ 0.0      & 0.90 $\pm$ 0.1                  \\ \hline
			\textbf{$\mathcal{D}_p$}                                  & 0.87 $\pm$ 0.0      & 0.92 $\pm$ 0.0        & 0.72 $\pm$ 0.1      & 0.73 $\pm$ 0.1                  \\ \hline
			\textbf{$\mathcal{D}_h$}                                  & 0.88 $\pm$ 0.0      & 0.93 $\pm$ 0.0         & 0.88 $\pm$ 0.0      & 0.93 $\pm$ 0.0         \\ \hline
		\end{tabular}\par\vspace{0.5em}
		(b) 
	\end{minipage}
	
\end{table}

 \begin{figure}[t] 
 	\centering
 	\includegraphics[width=0.9\textwidth]{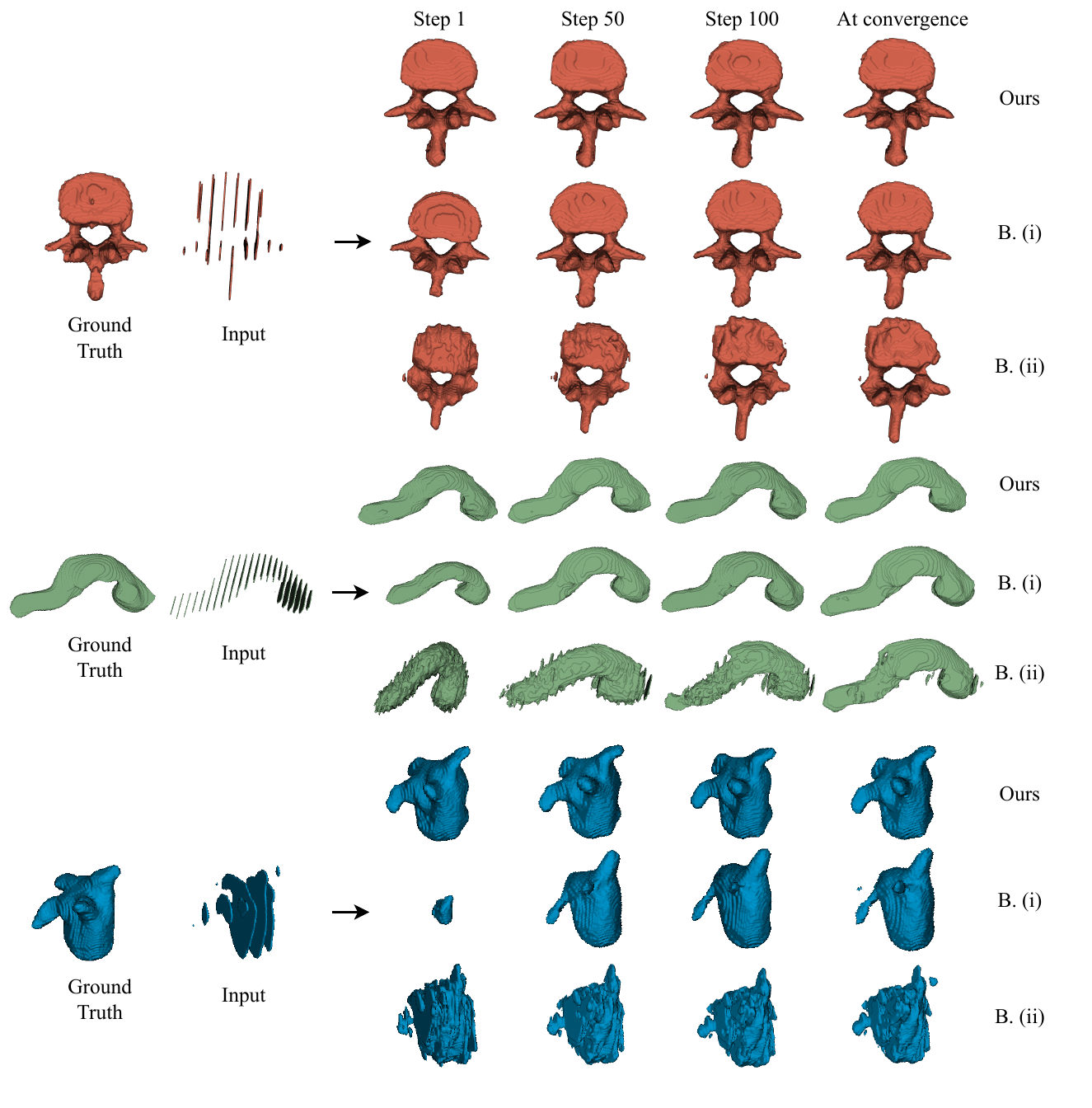}
 	\caption{Qualitative comparisons of reconstruction performance at different optimization steps and at convergence.}
 	\label{fig:alldata}
 \end{figure}
 
Qualitatively (Fig.~\ref{fig:vertebra}, \ref{fig:alldata}), we notice that after one optimization step both B.~(i) and B.~(ii) provide a shape in the training distribution which can be interpret as the prior learned by these models, that however is not yet specialized to that specific test sample. 

These results suggest that the meta-learned initialization $\theta_0^*$ holds significant advantages for the swift reconstruction of medical shapes, indicating its potential use for real-time applications. We emphasize that we also show this to be true in $\mathcal{D}_h$, which consists of few training samples, showcasing that our model is able to learn a strong initialization from limited training data.

As our focus is on time-constrained environments, we report results at one and 100 steps in the next evaluations, as they generally align closely with results obtained at convergence (Table~\ref{tab:steps}).

\subsection{Generalizing from different reconstruction tasks}

\subsubsection{Lower input resolution} \label{sec:resultOffset}
We reconstruct shapes in the sagittal direction setting a greater distance between the slices, i.e. $w = 16$. Comparisons with baselines are reported in Tables~\ref{tab:sagittal} (a). The Average Surface Distance (ASD) evaluation is reported in Table~\ref{tab:sagittalASD} in Appendix.
In this scenario, our model demonstrates quantitative superiority over other methods for just one optimization step. This underscores its capacity to swiftly specialize to a new shape, while being able to generalize when fewer observations are available. 
\subsubsection{Axial and Coronal Slices}
We additionally performed the experiments described in Sec.~\ref{sec:resultSagittal} in axial and coronal sampling directions. We achieve similar results to reconstruction from sagittal slices (Table~\ref{tab:sagittal} (b)), demonstrating our model can be used to quickly reconstruct across observations in different directions. Comparisons with the baselines are given in Table~\ref{tab:coronalaxial} in Appendix.

\begin{table}[t]
	\centering

		\caption{Average Dice score for transfer learning experiments at 1 ($DSC_1$) and 100 ($DSC_{100}$) optimization steps. The shape prior training uses the datasets in the top row e.g. $D_v$. Each pre-training is evaluated on other two shapes e.g. $D_p, D_h$.}

	\begin{tabular}{lcccccc}
		\hline
		\multicolumn{1}{c}{\textbf{}} & \multicolumn{2}{c}{\textbf{$D_v$}}                                                                                 & \multicolumn{2}{c}{\textbf{$D_p$}}                                                                                 & \multicolumn{2}{c}{\textbf{ $D_h$}}                                                                                   \\ \hline
		\multicolumn{1}{c}{}          & \begin{tabular}[c]{@{}c@{}}$DSC_{1}$\\ $D_p$\end{tabular} & \begin{tabular}[c]{@{}c@{}}$DSC_{100}$\\ $D_p$\end{tabular} & \begin{tabular}[c]{@{}c@{}}$DSC_1$ \\ $D_v$\end{tabular} & \begin{tabular}[c]{@{}c@{}}$DSC_{100}$ \\ $D_v$\end{tabular} & \begin{tabular}[c]{@{}c@{}}$DSC_{1}$ \\ $D_v$\end{tabular} & \begin{tabular}[c]{@{}c@{}}$DSC_{100}$ \\ $D_v$\end{tabular} \\ \hline
		B.~(i)                        & 0.01 $\pm$ 0.0                                            & 0.27 $\pm$ 0.1                                              & 0.01 $\pm$ 0.0                                           & 0.52 $\pm$ 0.2                                               & 0.01 $\pm$ 0.0                                             & 0.54 $\pm$ 0.1                                               \\ \hline
		B.~(ii)                       & 0.13 $\pm$ 0.0                                            & 0.75 $\pm$ 0.1                                              & 0.19 $\pm$ 0.0                                           & 0.75 $\pm$ 0.1                                               & 0.30 $\pm$ 0.0                                             & 0.66 $\pm$ 0.0                                              \\ \hline
		Ours                          & \textbf{0.82 $\pm$ 0.0}                                   & \textbf{0.91 $\pm$ 0.0}                                     & \textbf{0.70 $\pm$ 0.0}                                  & \textbf{0.80 $\pm$ 0.0}                                      & \textbf{0.50 $\pm$ 0.1}                                    & \textbf{0.70 $\pm$ 0.1}                                      \\ \hline
		\multicolumn{1}{c}{}          & 
		\begin{tabular}[c]{@{}c@{}}$D_h$\end{tabular} & \begin{tabular}[c]{@{}c@{}}$D_h$\end{tabular} & \begin{tabular}[c]{@{}c@{}}$D_h$\end{tabular} & \begin{tabular}[c]{@{}c@{}}$D_h$\end{tabular} & \begin{tabular}[c]{@{}c@{}} $D_p$\end{tabular} & \begin{tabular}[c]{@{}c@{}}$D_p$\end{tabular} \\ \hline
		B.~(i)                        & 0.30 $\pm$ 0.0                                            & 0.83 $\pm$ 0.0                                              & 0.01 $\pm$ 0.0                                           & 0.72 $\pm$ 0.1                                               & 0.01 $\pm$ 0.0                                             & 0.08 $\pm$ 0.1                                               \\ \hline
		B.~(ii)                       & 0.24 $\pm$ 0.1                                            & 0.80 $\pm$ 0.0                                              & 0.15 $\pm$ 0.0                                           & 0.64 $\pm$ 0.1                                               & 0.11 $\pm$ 0.1                                             & \textbf{0.71 $\pm$ 0.1}                                      \\ \hline
		Ours                          & \textbf{0.87 $\pm$ 0.0}                                   & \textbf{0.91 $\pm$ 0.0}                                     & \textbf{0.81 $\pm$ 0.0}                                  & \textbf{0.89 $\pm$ 0.0}                                      & \textbf{0.38 $\pm$ 0.2}                                    & 0.66 $\pm$ 0.2                                               \\ \hline
	\end{tabular} 	\label{tab:transferlearning}
\end{table}

\begin{figure}[t] 
	\centering
	\includegraphics[width=\textwidth]{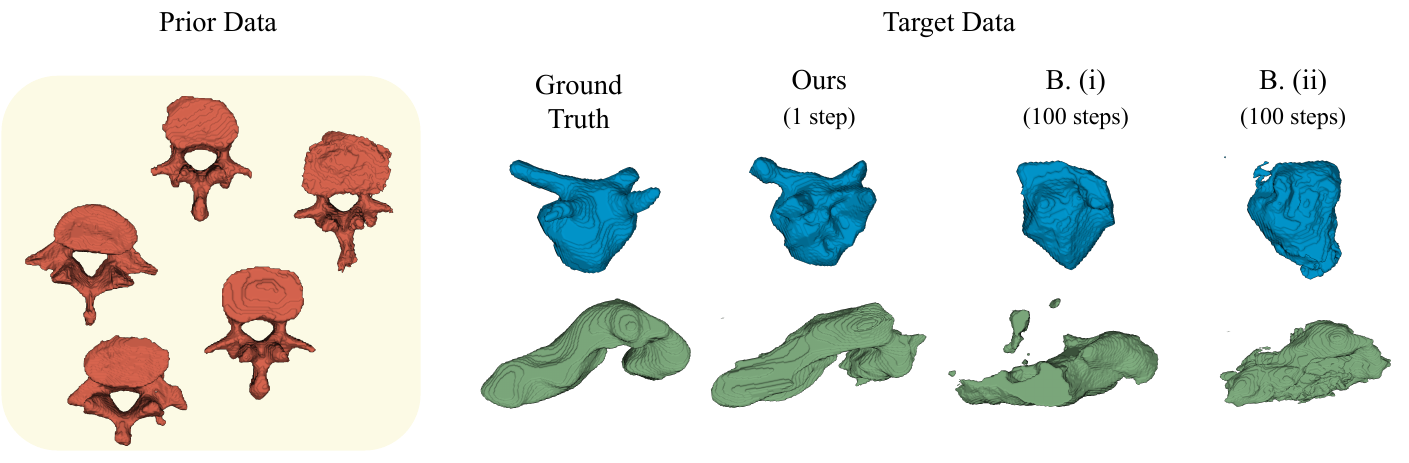}
	\caption{Transfer learning qualitative results example: reconstruction of heart and pancreas shapes from model trained on a different domain shape (vertebra).}
	\label{fig:ood}
\end{figure}

\subsubsection{Transfer Learning}
We investigate the capabilities of our model to generalize to target shape domains that differ from the prior domain.
Specifically, we evaluate each model trained on one prior shape e.g. vertebra, on other target domains e.g. pancreas and heart. The input configuration is the one described in Sec.~\ref{sec:resultSagittal}. Fig~\ref{fig:ood} visualizes examples of reconstructions from all the methods, when trained on vertebra and evaluated on heart and pancreas datasets. The meta-learning approach outperforms the baseline methods for all tested shapes, after both 1 and 100 optimization steps (Table~\ref{tab:transferlearning}). 
This suggests that our method is not limited to optimize representations in the learned latent space, demonstrating transferable capabilities to different shape domains. Therefore, this experiment show that our approach can be applied also when available target samples are insufficient for training, a common scenario in medical domain.

\section{Conclusion}
We present a meta-learning based implicit representation method to fast recover complete anatomical shapes from limited observations. 
With an extensive evaluation, we show that the meta-learned initialization is effective within this scope. Our method achieves faster reconstruction times compared to current state-of-the-art, maintaining high accuracy. Therefore, this approach could be valuable in a clinical setting, particularly for real-time applications such as surgical navigation.
Additionally, we demonstrate that our model can generalize across different tasks such as sparse slices with different orientations and spacings, without requiring a large amount of training data. Notably, our framework exhibits transferable abilities in reconstructing shapes from target domains different from the training domain, highlighting the potential for further exploration in this direction.
In future work, further improvement can be made including refining the meta-learning algorithm to improve reconstruction accuracy and implementing architectural enhancements to better align with the data manifold.

\section*{Appendix}

\begin{table}[h]
	\caption{Average $DSC$, number of iterations and corresponding time reached at convergence, for reconstructions from input segmentations with $w = 8$ along sagittal axis.}
	\centering
	\begin{tabular}{llccc}
		\cline{2-5}
		& \multicolumn{1}{c}{\textbf{}} & \textbf{$DSC_{conv}$}           & \textbf{$\#$ iters}      & \textbf{\begin{tabular}[c]{@{}c@{}}Rec. time (s)\end{tabular}} \\ \cline{2-5} 
		\textbf{$\mathcal{D}_v$} & B. (i)                        & \textbf{0.92 $\pm$ 0.02} & 143.2 $\pm$ 34.2         & \textbf{4.25 $\pm$ 1.0}                                          \\ \cline{2-5} 
		& B. (ii)                       & 0.88 $\pm$ 0.02          & 201.9 $\pm$ 37.4         & 6.97 $\pm$ 1.3                                                   \\ \cline{2-5} 
		& Ours                          & 0.91 $\pm$ 0.02          & \textbf{56.7 $\pm$ 18.2} & 7.29 $\pm$ 2.3                                                   \\ \cline{2-5} 
		\textbf{$\mathcal{D}_p$} & B. (i)                        & 0.94 $\pm$ 0.03          & 182.9 $\pm$ 43.8         & \textbf{5.40 $\pm$ 1.3}                                          \\ \cline{2-5} 
		& B. (ii)                       & \textbf{0.95 $\pm$ 0.01} & 255.1 $\pm$ 35.3         & 8.79 $\pm$ 1.2                                                   \\ \cline{2-5} 
		& Ours                          & 0.93 $\pm$ 0.02          & \textbf{97.3 $\pm$ 43.9} & 13.42 $\pm$ 6.1                                                  \\ \cline{2-5} 
		\textbf{$\mathcal{D}_h$} & B. (i)                        & \textbf{0.93 $\pm$ 0.02} & 154.6 $\pm$ 29.0         & \textbf{4.61 $\pm$ 0.9}                                          \\ \cline{2-5} 
		& B. (ii)                       & 0.78 $\pm$ 0.03          & 250.7 $\pm$ 43.7         & 8.62 $\pm$ 1.6                                                   \\ \cline{2-5} 
		& Ours                          & 0.91 $\pm$ 0.02          & \textbf{48.9 $\pm$ 24.3} & 6.25 $\pm$ 3.1                                                   \\ \cline{2-5} 
	\end{tabular}
	\label{tab:convergence}
\end{table}

\begin{table}
	\centering
	\caption{Average Surface Distance for vertebra ($\mathcal{D}_v$), pancreas ($\mathcal{D}_p$) and heart ($\mathcal{D}_h$) reconstructions from input segmentations with $w = 8$ (a) and $w = 16$ (b) along sagittal axis, after 1 ($DSC_1$) and 100 ($DSC_{100}$) optimization steps.}
\label{tab:sagittalASD}
	\begin{minipage}{0.45\textwidth}
		\centering
	\begin{tabular}{llcl} 
		\cline{2-4}
		& \multicolumn{1}{c}{\textbf{}} & $ASD_1$                           & \multicolumn{1}{c}{$ASD_{100}$}  \\ 
		\cline{2-4}
		\multicolumn{1}{c}{\textbf{$D_v$}} & B.~(i)                        & 3.09~$\pm$~1.25                   & \textbf{0.44~$\pm$ 0.13}         \\ 
		\cline{2-4}
		& B.~(ii)                       & 3.26~$\pm$~1.20                   & 1.22~$\pm$ 0.31                  \\ 
		\cline{2-4}
		& Ours                          & \textbf{0.71~$\pm$ 0.20}          & 0.51~$\pm$ 0.11                  \\ 
		\cline{2-4}
		\multicolumn{1}{c}{\textbf{$D_p$}} & B.~(i)                        & 5.74 $\pm$ 2.71                     & 0.39 $\pm$ 0.25                  \\ 
		\cline{2-4}
		& B.~(ii)                       & 10.43 $\pm$ 3.72                   & 3.29 $\pm$ 1.19                         \\ 
		\cline{2-4}
		& Ours                          & \textbf{0.79 $\pm$ 0.44} & \textbf{0.33 $\pm$ 0.13}       \\ 
		\cline{2-4}
		\multicolumn{1}{c}{\textbf{$D_h$}} & B.~(i)                        & 11.42 $\pm$ 1.46                        & \textbf{0.66 $\pm$ 0.20}         \\ 
		\cline{2-4}
		& B.~(ii)                       & 0.22 $\pm$ 0.11                   & 0.67 $\pm$ 0.05                  \\ 
		\cline{2-4}
		& Ours                          & \textbf{\textbf{1.40 $\pm$ 0.45}} & 0.99~$\pm$ 0.44                  \\
		\cline{2-4}

\end{tabular}\par\vspace{0.5em}
(a) 
\end{minipage}
	\begin{minipage}{0.45\textwidth}
	\centering
	\begin{tabular}{llcl} 
		\cline{2-4}
		& \multicolumn{1}{c}{\textbf{}} & $ASD_1$                  & \multicolumn{1}{c}{$ASD_{100}$}      \\ 
		\cline{2-4}
		\multicolumn{1}{c}{\textbf{$D_v$}} & B.~(i)                        & 3.10 $\pm$ 1.25       & \textbf{0.61 $\pm$ 0.21}                 \\ 
		\cline{2-4}
		& B.~(ii)                       & 3.28 $\pm$ 1.21       & \multicolumn{1}{c}{1.80 $\pm$ 0.49}  \\ 
		\cline{2-4}
		& Ours                          & \textbf{1.34 $\pm$ 0.33}  & 1.22~$\pm$ 0.26                      \\ 
		\cline{2-4}
		\multicolumn{1}{c}{\textbf{$D_p$}} & B.~(i)                        & 5.74~$\pm$~2.72          & \textbf{0.48~$\pm$ 0.26}             \\ 
		\cline{2-4}
		& B.~(ii)                       & 4.99 $\pm$ 1.96          & 2.45 $\pm$ 0.80                             \\ 
		\cline{2-4}
		& Ours                          & \textbf{0.93 $\pm$ 0.52} & 0.60~$\pm$ 0.27                      \\ 
		\cline{2-4}
		\multicolumn{1}{c}{\textbf{$D_h$}} & B.~(i)                        & 11.42 $\pm$ 1.46         & \textbf{0.71~$\pm$ 0.22}             \\ 
		\cline{2-4}
		& B.~(ii)                       & 10.90 $\pm$ 3.37         & 6.55 $\pm$ 1.70                      \\ 
		\cline{2-4}
		& Ours                          & \textbf{2.42~$\pm$ 0.93} & \multicolumn{1}{c}{2.24 $\pm$ 0.93}  \\
		\cline{2-4}
\end{tabular}\par\vspace{0.5em}
(b) 
\end{minipage}

\end{table}

\begin{table}[]
		\centering
		
\caption{Average DSC for vertebra ($\mathcal{D}_v$), pancreas ($\mathcal{D}_p$) and heart ($\mathcal{D}_h$) reconstructions from input segmentations with sampled slices with offset $w = 8$ along the coronal and axial axis, after 1 ($DSC_1$) and 100 ($DSC_{100}$) optimization steps.}
\label{tab:coronalaxial}
	\begin{tabular}{lllll}
		\hline                                                                                   
		\multicolumn{1}{c}{}                  & \multicolumn{2}{c}{\textit{\textbf{Coronal}}}              & \multicolumn{2}{c}{\textit{\textbf{Axial}}}                \\ \hline
		& \multicolumn{1}{c}{$DSC_1$} & \multicolumn{1}{c}{$DSC_{100}$} & \multicolumn{1}{c}{$DSC_1$} & \multicolumn{1}{c}{$DSC_{100}$} \\ \hline
		\multicolumn{1}{c}{\textbf{$\mathcal{D}_v$}} &                            &                               &                            &                               \\ \hline
		B.~(i)                          & 0.60 $\pm$ 0.1               & 0.90 $\pm$  0.0                 & 0.60 $\pm$ 0.1               & \textbf{0.92 $\pm$ 0.0}         \\ \hline
		B.~(ii)                         & 0.63 +. 0.1               & 0.80 $\pm$ 0.1                  & 0.64 $\pm$ 0.1               & 0.84 $\pm$ 0.0                  \\ \hline
		Ours                                  & \textbf{0.89 $\pm$ 0.0}      & \textbf{0.91 $\pm$ 0.1}         & \textbf{0.88 $\pm$ 0.0}      & 0.90 $\pm$ 0.1                  \\ \hline
		\multicolumn{1}{c}{\textbf{$\mathcal{D}_p$}} &                            &                               &                            &                               \\ \hline
		B.~(i)                          & 0.35 $\pm$ 0.1               & \textbf{0.92 $\pm$ 0.0}         & 0.34 $\pm$ 0.1               & \textbf{0.83 $\pm$ 0.1}         \\ \hline
		B.~(ii)                         & 0.43 $\pm$ 0.1               & 0.68 $\pm$ 0.1                  & 0.43 $\pm$ 0.1               & 0.58 $\pm$ 0.1                  \\ \hline
		Ours                                  & \textbf{0.87 $\pm$ 0.0}      & \textbf{0.92 $\pm$ 0.0}         & \textbf{0.72 $\pm$ 0.1}      & 0.73 $\pm$ 0.1                  \\ \hline
		\multicolumn{1}{c}{\textbf{$\mathcal{D}_h$}}    &                            &                               &                            &                               \\ \hline
		B.~(i)                          & 0.07 $\pm$ 0.0               & \textbf{0.93 $\pm$ 0.0}         & 0.07 $\pm$ 0.0               & \textbf{0.93 $\pm$ 0.0}         \\ \hline
		B.~(ii)                         & 0.22 $\pm$ 0.1               & 0.81 $\pm$ 0.0                  & 0.22 $\pm$ 0.1               & 0.81 $\pm$ 0.0                  \\ \hline
		Ours                                  & \textbf{0.88 $\pm$ 0.0}      & \textbf{0.93 $\pm$ 0.0}         & \textbf{0.88 $\pm$ 0.0}      & \textbf{0.93 $\pm$ 0.0}         \\ \hline
	\end{tabular}
\end{table}

%
%
%
\newpage
\bibliographystyle{splncs04}
\bibliography{biblio}

\end{document}